\documentclass[pss]{wiley2sp} % provides pss two-column style
\usepackage{amsmath}

\usepackage{times}
\usepackage{latexsym}
\usepackage{dcolumn}
\usepackage{amsfonts,amssymb}
\usepackage{graphicx,epsfig}
\usepackage{psfrag}

\begin{document}

\newcommand{\be}{\begin{equation}}
\newcommand{\ee}{\end{equation}}
\newcommand{\bearr}{\begin{eqnarray}}
\newcommand{\eearr}{\end{eqnarray}}
\newcommand{\nn}{\nonumber}
\newcommand{\vk}{\vec k}
\newcommand{\vp}{\vec p}
\newcommand{\vq}{\vec q}
\newcommand{\vkp}{\vec {k'}}
\newcommand{\vpp}{\vec {p'}}
\newcommand{\vqp}{\vec {q'}}
\newcommand{\up}{\uparrow}
\newcommand{\down}{\downarrow}
\newcommand{\fns}{\footnotesize}
\newcommand{\ns}{\normalsize}
\newcommand{\cdag}{c^{\dagger}}

\title{Neutral Triplet Collective Mode in Doped Graphene}
\titlerunning{Magnetic Excitations in graphene}

\author{
M. Ebrahimkhas	\textsuperscript{\textsf{\bfseries 1}}
S. A. Jafari	\textsuperscript{\Ast,\textsf{\bfseries 2,3,4}}
G. Baskaran	\textsuperscript{\textsf{\bfseries 5}} 
}

\authorrunning{M. Ebrahimkhas et al.}

\mail{e-mail
  \textsf{jafari@sharif.edu}, Phone:
  +98-21-66164524, Fax: +98-21-66022711}

\institute{
\textsuperscript{1}\,Department of Science, Tarbiat Modares University, Tehran 14115-175, Iran\\
\textsuperscript{2}\,Department of Physics, Sharif University of Technology, Tehran 11155-9161, Iran\\
\textsuperscript{3}\,Department of Physics, Isfahan University of Technology, Isfahan 84156-83111, Iran\\
\textsuperscript{4}\,School of Physics, Institute for Research in Fundamental Sciences (IPM), Tehran 19395-5531, Iran\\
\textsuperscript{5}\,Institute of Mathematical Sciences, Chennai 600113, India
}

\received{}
\published{}

\abstract{%
% Usage: \abstcol{<left column>}{<right column>}
\abstcol{%
Particle-hole continuum in Dirac sea of graphene has a unique window underneath, which 
provides a unique opportunity for emergence of a pole in the susceptibility of the {\em triplet} particle-hole
channel in the entire Brillouin zone (BZ). Here we use random phase approximation (RPA) 
to study such collective mode at zero 
temperature, in a single layer of doped graphene. We find that due to the chiral
nature of one-particle states, in undoped graphene, the wave function overlap factors 
do not lead to qualitative differences, while in doped graphene they will kill 
small momentum part of the branch of magnetic excitations by pushing it to touch 
the lower part of the continuum. The pole corresponding to magnetic excitations
survives for for larger momenta in the BZ.
}
{}
}

\titlefigure[height=3.1cm]{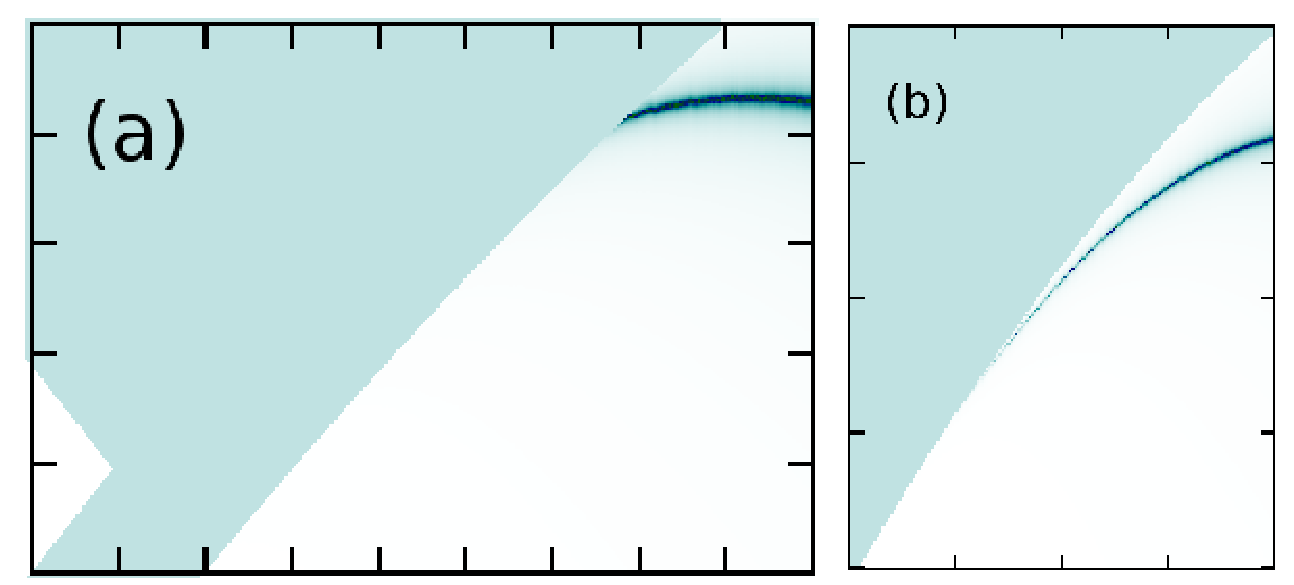}
\titlefigurecaption{(Color online) Continuum of the inter-band and 
intra-band particle-hole excitations is the color-filled region. 
The dark line is intensity plot corresponding to the magnetic 
excitations in (a) doped, and (b) undoped graphene.
}

%\titlefigure[height=3.1cm]{empty2w}
%\titlefigurecaption{ }

\maketitle

\section{Introduction}
Graphene, a single atomic layer of graphite was the first realization of a two dimensional 
elemental metallic structure~\cite{Novoselov}. The salient feature in the electronic spectrum 
of graphene, which distinguishes it from other materials, is the presence of Dirac cones in 
its dispersion~\cite{NetoRMP,Rotenberg,Bostwick}, which provides a laboratory to test 
relativistic type phenomena in the $\sim 1$ eV energy scale. 
The cone-like spectrum protects the system against impurity scattering~\cite{Amini2009}, as 
well as many-body interactions~\cite{Jafari2009}, both in normal~\cite{DasSarmaBornApprox,Qaiumzadeh} 
and superconducting phases~\cite{Garg}. Because of the robust mass-less cone-like dispersion 
with a large Fermi velocity in graphene, it is easier to  observe phenomena such as quantum 
Hall effect~\cite{Jiang}. In standard 2D electron gas systems, this effect appears only at 
low temperatures and for very pure samples, while in graphene it can be observed at ambient 
temperature.  The high mobility of careers in graphene at room temperature which is not 
appreciably different from its value at the liquid-helium temperature~\cite{Jiang,Berger} 
is another promising property of graphene for device applications.
Stacking the graphene with further layers produces graphene 
multi-layers, such as bilayer, etc. For few layers, due to quantum size effect,
the cone like dispersion is replaced by other types of chiral dispersions~\cite{NetoRMP}.
However, when the number of stacks is large enough ($\approx 10$) to
approach the bulk limit of graphite, the cone like dispersion
is again recovered~\cite{Wallace}, except for small electron-hole pockets
at very low-energies $\sim 40$ meV. Therefore the phenomena driven by the
Dirac nature of careers is common in graphene and 
graphite~\cite{ARPESgraphite,LiAndrei,Orlita,BaskaranJafari,JafariBaskaran,Ebrahimkhas,kopel}.
These properties should also be shared with the recently fabricated multi-layer epitaxial 
graphene~\cite{Conard}.

After the pioneering work of Wallace~\cite{Wallace} on the tight binding band picture 
for the electronic structure of pure and undoped graphene and graphite, it has become 
popular to take into account the effects of disorder, interactions and doping on top 
of a band picture~\cite{NetoRMP}. Nevertheless, there exists an alternate quantum 
chemical approach to the electronic properties of graphene: More than half a century ago, 
Pauling argued that the ground state of graphene can be described as a natural extension 
of the resonating valence bond (RVB) state of benzene~\cite{PaulingBook,BaskaranMgB2,Doniach}, 
but he totally ignored unbound polar (charge fluctuation) configurations. This overemphasize on 
neutral configuration makes graphene a Mott insulator. But graphene and graphite are both 
semi metals in reality. Since then both band aspect and Pauling's singlet correlations 
have been argued to be present in graphene~\cite{Azadi} and two important consequences have been brought 
theoretically. One of them is existence of 
gap-less neutral triplet bosonic mode~\cite{BaskaranJafari,JafariBaskaran} for neutral ({\em i.e.} undoped) graphene, 
which was interpreted as a two-spinon bound excited state above a long-range RVB ground state~\cite{Noorbakhsh}.
The other is a suggestion of high temperature superconductivity in doped graphene~\cite{Shenoy,Sahebsara} 
and other exotic superconducting states~\cite{CastroNetoSC,Honerkamp}. 
An insulating RVB state is found to be stabilized~\cite{Noorbakhsh} at least in the 
Mott insulating side of the phase diagram~\cite{Jafari2009}. Moreover, lattice gauge
theory simulation of $2+1$ dimensional QED predicts the  critical value of 
the "fine structure" constant in graphene can be crossed in suspended graphene~\cite{Drut2009}. 
In this scenario, the ground state of graphene in vacuum is expected to be a Mott insulator.
Recent investigation of finite clusters of $sp^2$ bonded systems quantum Monte Carlo 
methods~\cite{Azadi} has revealed substantial RVB correlations
in undoped graphene. Moreover, two collective spin and charge excitations
in these system have been found~\cite{Kaveh} which are argued to be naturally 
understandable in terms of an underlying RVB ground state~\cite{Kaveh}.
As is detailed in the following, the collective spin state in the undoped case 
can alternatively be understood even from a weak coupling side within 
a simple random phase approximation~\cite{BaskaranJafari}.
In this work we would investigate the fate of such triplet
collective excitations in presence of doping.

This paper is organized as follows:
We start by discussing the nature of free particle-hole excitations
in doped and undoped graphene. Next we introduce the model and summarize 
the RPA formulation. In the small $q$ limit where the linearized
Dirac theory around the K points is valid, closed form expressions for the susceptibility
can be obtained~\cite{Wunsch,HwangSarma}. Otherwise we report numerical results.
We end the paper with a summary and discussion.

\section{Nature of free particle-hole excitations}
 %figure  
\begin{figure}[b]
\begin{center}
\includegraphics[width=8cm,height=5cm,angle=0]{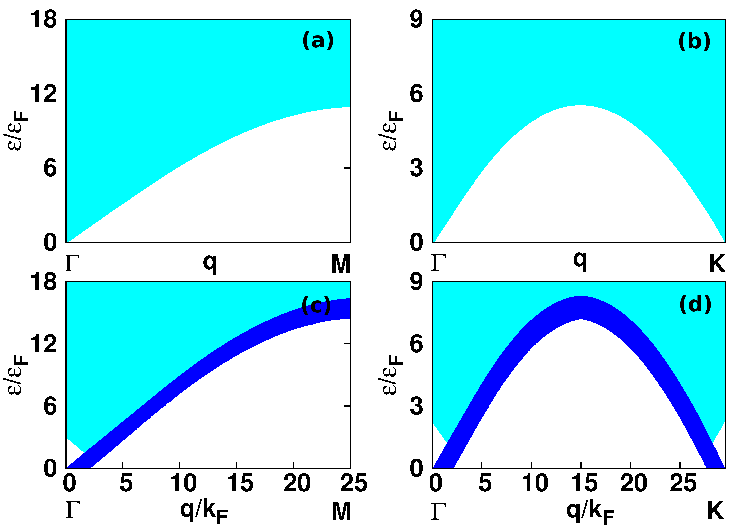}
\includegraphics[width=7cm,height=4cm,angle=0]{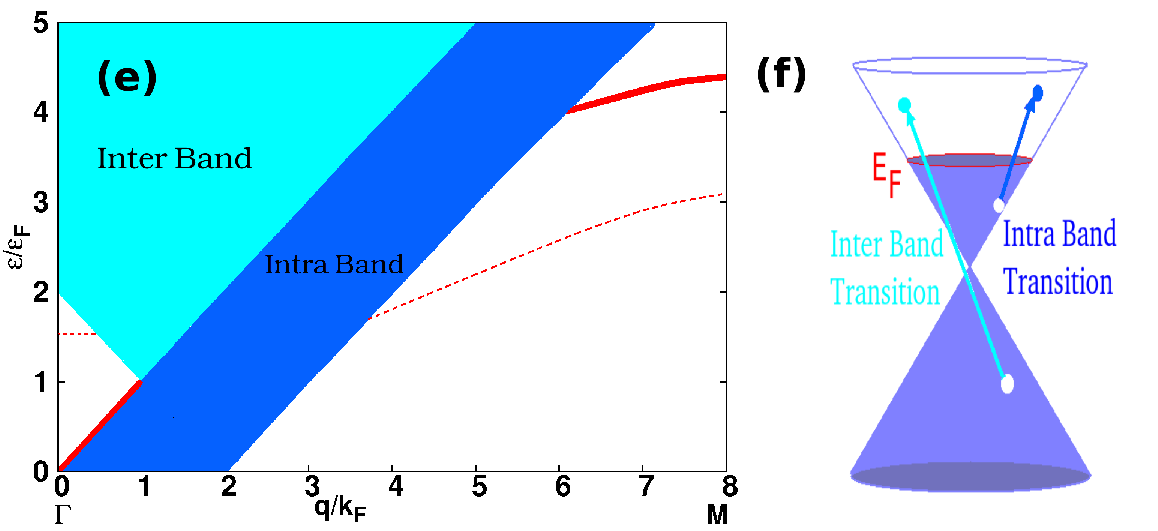}
\vspace{-4mm}
\caption{(Color online) 
Particle-hole continuum of graphene band structure  
along important directions of the Brillouin zone. (a) and (b) denote
the PHC for undoped graphene, (c) and (d) correspond to doped graphene. 
For better comparison, the energies of both doped and undoped cases are reported 
in units of $\varepsilon_F$.
Panel (e) is the same as (c) enlarged for clarity, 
which also schematically shows the neutral spin-1 collective mode branch 
with (without) overlap factors by solid (dashed) red line.
Panel (f) schematically depicts the cone-like band
picture with two possible sets of inter-band and intra-band processes shown 
by cyan and blue arrows, respectively.
}
\label{PHC.fig}
\end{center}
\vspace{-4mm}
\end{figure}

\begin{figure}[tb]
\begin{center}
\includegraphics[width=5cm,height=4.2cm,angle=0]{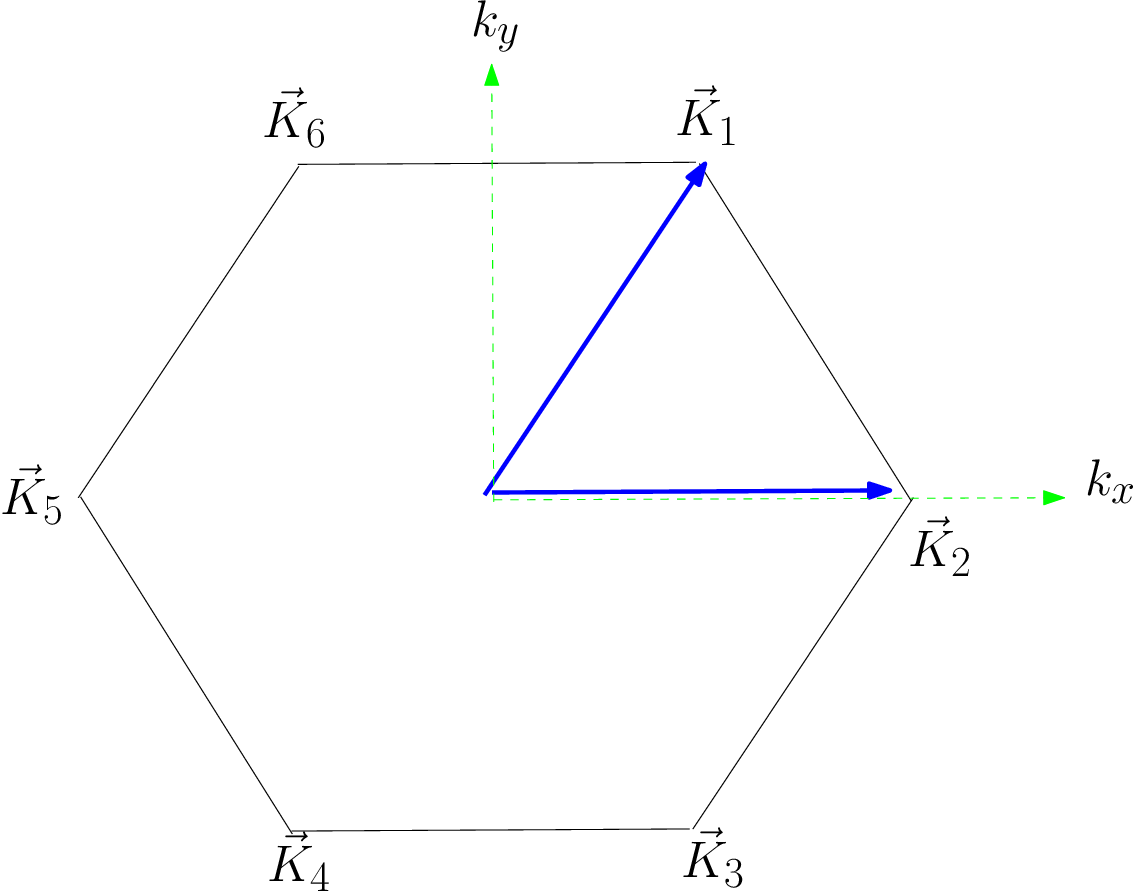}
\caption{(Color online) The BZ of graphene with inequivalent Dirac cones 
at $\vec K_1, \vec K_2$, etc. Only two of the cones are independent. 
Those with even (odd) indices are equivalent to each other by periodicity in 
the reciprocal space. 
}   
\label{BZ.fig}
\end{center}
\end{figure}

  The single-particle portion of the excitation spectrum in graphene is very well
described by a 2+1 dimensional Dirac theory. The interesting question here is, 
what are possible collective excitations arising from many-body effects when
one approaches the problem from weak coupling limit?
In doped graphene, a plasmon branch with square root dispersion has been
found in graphene~\cite{SeyllerHREELS}. 
More interesting many-body effects such as plasmaron can also be expected
in doped graphene~\cite{AsgariScience}. The plasmon excitations in the above works
is a collective excitation in {\em singlet} particle-hole channel. Within
simple random phase approximation (RPA), there can be no singlet 
bonding collective mode in undoped graphene, although going beyond RPA 
another branch of singlet collective excitations has been predicted
in undoped graphene~\cite{Ganga}. In this work we focus on the {\em triplet} channel
of the {\em doped} graphene. Can there be any interesting collective in this channel?
We have previously shown that even in undoped graphene, triplet channel
admits a collective branch of excitations~\cite{BaskaranJafari}. 
Doping of graphene can be achieved by applying appropriate gate voltage~\cite{NetoRMP}, or
other methods such as chemical doping\cite{Giovannetti}.
The bi-partite nature of honeycomb lattice
implies that the nearest neighbor tight-binding Hamiltonian considered
in our model must be particle-hole symmetric. Hence doping with electrons
and holes are treated on the same footing. Therefore here we focus
only on the case of electron doping.
As can be seen in Fig.~\ref{PHC.fig}~(f), doping with electrons have two effects: One is to Pauli
block some of the inter-band transitions which depletes the region of momenta
around the $\Gamma$ and $K$ points, compared to undoped case. Second is to add a new
2D like portion to the PHC corresponding to intra-band particle-hole
excitations denoted by blue in panel (e).
Therefore, populating the conduction band of the Dirac cone with electrons
modifies the particle-hole continuum (PHC) by adding a small
2D-like portion and creating small triangular windows around the 
$\Gamma$ and $K$ points of Brillouin zone, shown in Figure~\ref{PHC.fig}~(c,d,e)
by dark blue. 
The particle-hole continuum around the $K$ point is associated with inter-cone
scattering processes corresponding to various values of $\vq$ around  $\vec Q_{i}$ in Figure~\ref{BZ.fig}.
In the triangular window corresponding to small momenta around $\Gamma$ point, 
the triplet branch of collective excitations (solid red line) will actually touch the intra-band 
portion of the PHC tangentially, and will not be well-defined branch. 
For larger momenta it will emerge again in the other side below the whole PHC.

%figure 

\section{Formulation of the problem}
Unlike plasmon (singlet) excitations, for which the long-range part of the Coulomb 
interaction is essential, since here we are interested in collective excitations
in triplet (spin-flip) channel, we only need to consider the short range part of the 
interaction. 
A recent ab-initio calculation substantiates this assumption,
according the which even the screened value for the short range (Hubbard) 
part of the interaction in graphene was obtained to be about $9$ eV~\cite{Wehling}.
The next neighbor Coulomb parameter ($U_{01}$) in Ref.~\cite{Wehling} was obtained to 
be $\sim 4$ eV. As far as the formation of a triplet low-energy state is 
concerned, it has been shown that inclusion of longer range part of
the interactions does not lead to qualitative change in the dispersion of
spin-1 collective excitations~\cite{JafariBaskaran}. The next neighbor Coulomb
parameter may have its own interesting effects in creating many-body states
similar to light-excitons~\cite{Hafez}, while the triplet excitations considered
in our present work can be considered as analogue of dark excitons. Such states
maybe combined in doped situations to give rise to more complicated objects, 
such as recently observed trions in hole-doped carbon nano-tubes~\cite{Matsunaga}.

Hence the model we use is the Hubbard model defined as,
\begin{equation}
   H=-t\sum_{\langle i,j\rangle,\sigma} 
   \left( c^\dagger_{i\sigma}c_{j\sigma}+{\rm h.c.}\right)
   +U\sum_j n_{j\down} n_{j\up},
\end{equation}
where $i,j$ denote sites of a honeycomb lattice, and $\sigma$
stands for spin of electrons. 
The spectrum of the band limit, $U=0$, is given by
$\epsilon_{\vk, \alpha}=\pm\epsilon_{\vk}$ corresponding to $\alpha=+1,-1$
for conduction and valence bands, respectively, and,
\be
   \epsilon_{\vk}=t\sqrt{1+\cos(\sqrt{3}k_{y}/2)\cos(k_{x}/2)+4\cos^{2}(k_{x}/2)},
   \label{band.eqn}
\ee
where $t\sim 2.8$ eV is the hopping amplitude. 
The lattice parameter $a$ is taken as carbon-carbon distance.
In this model, 
$U$ is on-site Coulomb repulsion. Although the bare value
of $U$ in graphene is $\sim 4t-5t$, but within the RPA one should 
not increase $U$ above $U_c\sim 2.23t$~\cite{JafariBaskaran}. 
The underestimation of $U_c$ is a known artifact of RPA, as in the
sense of Hubbard-Stratonivich transformation, the RPA approximation 
belongs to the family of mean field approximations~\cite{NagaosaBook}.
Therefore to be consistent in applying the RPA approximation,
we restrict ourselves to values of $U\sim 2t$.

We implement the RPA approximation in the triplet particle-hole
channel, which is given by~\cite{BickersScalapino,EOM},
\be
   \chi^{\rm RPA}_{\rm triplet}(\vec q,\omega)=
   \frac{\chi^{(0)}(\vec q,\omega)}{1-U\chi^{(0)}(\vec q,\omega)}.
   \label{rpatriplet.eqn}
\ee
Note that the sign of $U$ for triplet and singlet channels is 
different~\cite{BickersScalapino}. Hence, when the above triplet susceptibility diverges,
the contribution of the singlet channel to the total susceptibility
will remain finite. 
The retarded bare susceptibility $\chi^{(0)}$ is given by the standard
particle-hole form,
\bearr
   \chi^{(0)}(\vec{q},\omega)=
   \label{bare-polar.eqn}
   \frac{1}{N} \sum_{\vec{k}\alpha,\alpha'} 
   \frac{\left(f^{\alpha'}_{\vec{k}+\vec{q}}-f^{\alpha}_{\vec{k}}\right) F^{\alpha,\alpha'}(\vec k, \vec k + \vec q)}
   {\hbar\omega-(\epsilon_{{\vec{k}+\vec{q}},\alpha'}-\epsilon_{\vec{k},\alpha})+i0^{+}},
   \label{chi0.eqn}
\eearr
where $N$ is the number of unit cells and $\alpha ,\alpha'= \pm 1$ stand for 
conduction and valence bands, $f^{\alpha}_{\vk}$ is the Fermi distribution function, which determines 
the occupation of the state characterized with quantum labels $(\vk,\alpha)$
and energy $\epsilon_{\vk, \alpha}$, and wave function overlap factors are given by,
\be
   F^{\alpha,\alpha'}(\vk ,\vk+\vq)=\left(1+{\alpha\alpha'}\cos(\theta_{\vec k}-\theta_{\vec k+\vec q})\right)/2,
   \label{overlap.eqn}
\ee
where $\zeta_{\vk}\equiv e^{i\theta_{\vec k}}\equiv \phi(\vk)/|\phi(\vk)|$, 
$\phi(\vk)=-t\sum_{i=1}^{3} e^{i\vec \delta_{i}.\vk}$,
$\vec \delta_i$ is nearest neighbor vectors on the honeycomb lattice. 
These factors are simply due to the change of basis from 
two sub-lattice (A,B) into the physically transparent basis of conduction and
valence states ($\alpha$=+,-). 
The scattering matrix elements between chiral states $(\vec k,\alpha)$ and
$(\vec k',\alpha')$ of the cone-like dispersion in graphene are given
by $\langle \vec k',\alpha'|V|\vec k,\alpha\rangle=
\tilde V(\vec k-\vec k')\left(1+\alpha\alpha' e^{i\theta_{\vec k}-i\theta_{\vec k'}} \right)/2 $,
where $\tilde V$ is the Fourier transform
of the scattering potential. When the above phase factors are inserted into
particle-hole bubble diagrams, give rise to the overlap factor in the free
particle-hole propagator, Eq.~(\ref{overlap.eqn}).

Naive construction of  RPA like 
series in terms of density fluctuation bubbles gives rise to a second order 
equation in $U$, which does not have any solution in the triplet 
channel~\cite{PeresComment}. However, explicit construction of triplet 
operators along with arguments based on renormalization amounts instead 
of a second order equation, to two first order sets of equations, each one of 
which will be of the type $1-U\chi^{(0)}(\vq,\omega)=0$. One of these equations
admits solutions at finite values of interaction strength $U$~\cite{EOM}.
The valley degeneracy will be taken into account when comparing
numerical results in the hexagonal BZ with that of a linearized cone model
in a circular BZ of the same area. Also the spin degeneracy of $2$ appears
as a $3/2$ factors multiplying the whole $\chi^{\rm RPA}_{\rm triplet}$,
and another factor of $1/2$ multiplying $\chi^{\rm RPA}_{\rm singlet}$~\cite{BickersScalapino}.

\section{Results}
At $T  = 0$, and for electron doping case corresponding to $\mu > 0$, 
conduction band is partially occupied; i.e. from Dirac point to the Fermi level. 
So there are two types of particle-hole excitations: 
(i) {\em intra-band transition} corresponding to $\alpha, \alpha'=+1$.
(ii) {\em inter-band transitions} corresponding to $\alpha(\alpha')=-1(+1)$ in the above summation. 
The PHC corresponding to the above processes has been depicted in Fig.~\ref{PHC.fig},
and corresponds to regions in $\omega-\vq$ space where the imaginary part of
$\chi^{(0)}(\vq,\omega)$ takes on non-zero values.
The numerical calculation of $\chi^{(0)}(\vec q,\omega)$ for arbitrary $\vec q$
in the BZ is straightforward. However, for the low-energy part of 
the spectrum where the Dirac dispersion $\epsilon_{\vec k}=\hbar v_F|\vec k|$
governs the kinetic energy, one can obtain closed form formulae
for bare susceptibility~\cite{Wunsch,HwangSarma,Zeigler}.
Possible zeros of the denominator in Eq.~(\ref{rpatriplet.eqn}) occur
in regions where imaginary (dissipative) part is identically zero,
\be 
   \Re\chi^{(0)}(\vec{q},\omega)=\frac{1}{U},~~~~~~~~~~
   \Im \chi^{(0)}(\vec{q},\omega)=0.
   \label{collective-mode.eqn}
\ee 
The second equation above means that the solution must be outside
the PHC. Moreover in the first equation above, the right hand side
is positive, and so should be $\chi^{(0)}(\vq,\omega)$. But as can be
clearly seen from Eq.~\eqref{chi0.eqn}, the non-interacting susceptibility
can be positive only for $\hbar\omega < (\epsilon_{{\vec{k}+\vec{q}},+}-\epsilon_{\vec{k},-})$,
which actually defines the empty region below the PHC in Fig.~\ref{PHC.fig}.

\subsection{Undoped graphene}
In the case of undoped graphene, the calculation of $\chi^{(0)}$ defined in
Eq.~\eqref{chi0.eqn} gives the following result~\cite{Wunsch,HwangSarma}:
\begin{equation}
   \Im\chi^{(0)}(\vq, \omega) =\frac{q^{2}}{16}
   \frac{\theta(\omega-v_F q)}{\sqrt{\omega^2-v_F^2 q^2 }},
   \label{polar-undoped.eqn}
\end{equation}
which is surprisingly identical to the result obtained in Ref.~\cite{BaskaranJafari}
(Note that to compare to Ref.~\cite{BaskaranJafari} we have to set $g=1$).
Although in our earlier work~\cite{BaskaranJafari} on undoped graphene, we did
not take the overlap factors~\eqref{overlap.eqn} into account, we obtained
the same result. The first question in undoped graphene, neglect of the
wave-function overlap factors does not lead to a different result?
One qualitative way to understand this point is that, due to chiral
nature of electronic states in conduction and valence bands of the
Dirac cone, for particle-particle scattering in the conduction band 
(corresponding to $\alpha=\alpha'=1$) as well as for hole-hole 
scattering in the valence band (corresponding to $\alpha=\alpha'=-1$)
the back-scattering is diminished due to the overlap factors Eq.~\eqref{overlap.eqn},
as $1+{\alpha\alpha'}\cos(\theta_{\vec k}-\theta_{\vec k+\vec q})$ will be
zero for $\theta_{\vec k}-\theta_{\vec k+\vec q}=\pi$ when $\alpha\alpha'=1$.
Similarly, when $\alpha\alpha'=-1$, i.e.  for the particle-hole scattering the
forward scattering (corresponding to $\theta_{\vec k}-\theta_{\vec k+\vec q}=0$)
will be diminished. Therefore in the particle-hole channel, the back-scattering
contributes dominantly to the non-interacting susceptibility. 
Indeed a 1D like (inverse square root) behavior of the density of particle-hole
states is a result of such confinement of scattering to a line by enhancement
of back-scattering in the particle-hole channel. 
Using Eq.~(\ref{polar-undoped.eqn}) to solve Eq.~(\ref{collective-mode.eqn}) in
$q\to 0$ limits gives
\begin{equation}
   \omega(\vq)=v_F q-\frac{U^2}{32v_F }q^3,
   \label{s1-undoped.eqn}
\end{equation}
which is valid for a model of single Dirac cone.
%figure 
\begin{figure}[h]
\begin{center}
\vspace{-5mm}
\includegraphics[width=9cm,height=6cm,angle=0]{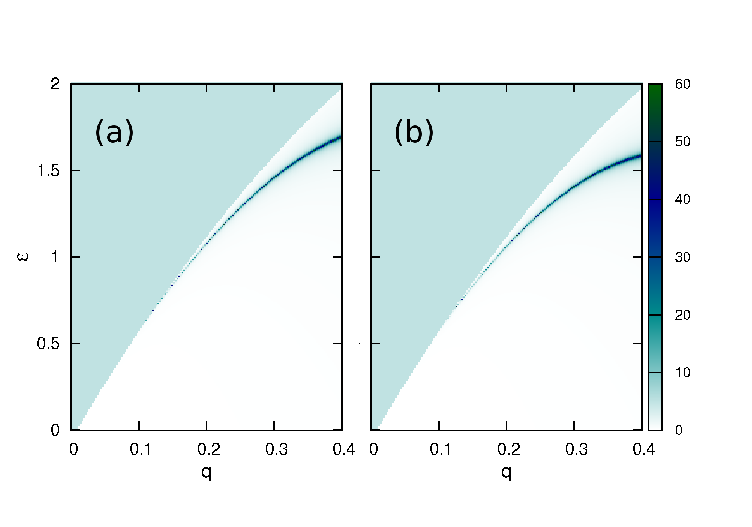}
\vspace{-8mm}
\caption{(Color online) Contour plot for $\Re\chi^{\rm RPA}_{\rm triplet}$ 
in $\Gamma \rightarrow K$ direction for undoped graphene, using the full energy 
band dispersion. Different panels correspond to (a) $U=1.8t$, (b) $U=2.0t$.
}
\label{undoped-overlap.fig}
\end{center}
\vspace{-4mm}
\end{figure}
Fig.~\ref{undoped-overlap.fig} shows the intensity plot for the 
dispersion of the triplet RPA susceptibility where the divergence
is clearly marked by intense line below the PHC in undoped graphene.
This result is obtained without linearization of the spectrum,
and hence effectively takes into account the presence of other cone 
as well. To compare results of a model with a single Dirac cone,
with those obtained from the the full tight-binding band dispersion, Eq.~(\ref{band.eqn}), 
which involves numerical integration of Eq.~(\ref{chi0.eqn}),
we have to keep the following point in mind: The values of $U$ in the 
linearized theory must be scaled in by the bandwidth $k_cv_F$
before comparing to values of $U/t$ in the tight-binding theory.
Moreover, the assumption of only one cone in a circular BZ, 
underestimated the particle-hole processes occurring at each cone 
by a factor of $2$. This leads to the scaling $U\to 0.4548 U$ in the
numerical results before comparing them to the analytic ones. 
With this point in mind, the result of 
numeric calculation is shown in Figure~\ref{undoped-overlap.fig}. As can be seen 
in the figure, taking into account the overlap factors for undoped graphene
again gives a dispersive spin-1 collective mode in the particle-hole channel,
in agreement with Ref.~\cite{BaskaranJafari}.

\subsection{Doped graphene}
Now let us concentrate on the case of doped graphene and calculate $\chi^{(0)}$
for $\mu>0$. The overlap factor $F^{+,-}$ which approached to $1$ in the limit of 
$\hbar\omega\to v_F|\vec q|\to 0$ was associated with the
inter-band processes which are relevant to undoped graphene.
However, in the case of doped graphene, inter-band processes at low momenta
are Pauli-blocked, and instead a new portion denoted by dark blue in Fig.~\ref{PHC.fig}
will be added to the continuum. Relevant to this portion are the intra-band 
overlap factors, $F^{+,+}$ for the electron doped case. In this case 
by Eq.~(\ref{overlap.eqn}) the back-scattered particle-hole pairs which
were responsible for the formation of a triplet bound state would actually
give zero. Because the condition $\cos(\theta_{\vec k}-\theta_{\vec k'})\to -1$ implies
$F^{+,+}\to 0$. Therefore in doped graphene, the phase space required for 
the formation of particle-hole bound state in triplet channels is 
diminished by the wave function overlap factors. Hence unlike the undoped case, 
in doped graphene we expect these overlap factors to play very crucial role at least
for momentum transfer around the $\Gamma$ point where the linearized
Dirac theory applies.

\begin{figure}[t]
\begin{center}
\includegraphics[width=5cm,height=7cm,angle=-90]{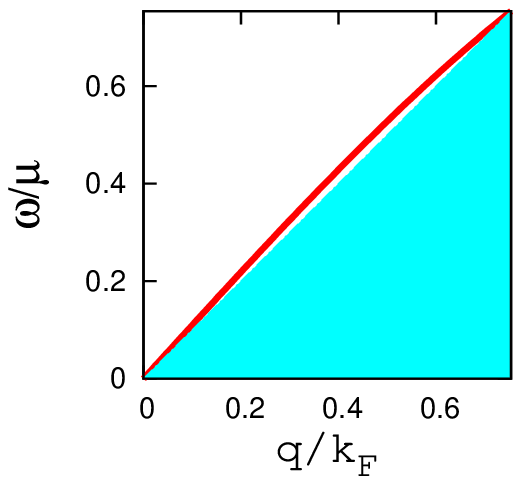}
\vspace{-3mm}
\caption{ (Color online) Color-filled region denotes PHC due to intra-band
particle-hole excitations. Dispersion relation Eq.~(\ref{s1-doped.eqn}) for $\gamma=0.1$.
In the limit $\mu\to 0$, we have $\gamma\to 0$ and spin-1 branch becomes tangent to 
the PHC edge. 
}   
\label{root.fig}
\end{center}
\end{figure}

Let us see this more formally in terms of analytic expressions.
The integral in Eq.~(\ref{chi0.eqn}) for the linearized 
Dirac cone theory can be calculated~\cite{Wunsch,HwangSarma} 
which in the $\vq\to 0$ limit becomes,
%to give:
%\bearr 
%   &&\Re\chi^{(0)}(\vq, \omega) =-\frac{g\mu}{2\pi\hbar^2 v^{2}_F}+
%    \frac{g\mu }{16\pi\hbar^2 v^2_F}\frac{x^2}{\sqrt{z^2-x^2}}\times\\ 
%   &&\left[-\left(\frac{2-z}{x}\right)\sqrt{\left(\frac{2-z}{x}\right)^2-1}+
%   \ln\left(\frac{2-z}{x}+\sqrt{\left(\frac{2-z}{x}\right)^2-1}\right)\right.\nn \\
%   &&\left.+\left(\frac{2+z}{x}\right)\sqrt{\left(\frac{2+z}{x}\right)^2-1}-
%   \ln\left(\frac{2+z}{x}+\sqrt{\left(\frac{2+z}{x}\right)^2-1}\right)\right],\nn
%\eearr
\bearr 
   &&\Re\chi^{(0)}(\vq, \omega) = -\frac{g\mu}{2\pi\hbar^2 v^{2}_F}\\
   &&+\frac{g\mu}{16\pi\hbar^2 v^2_F}
   \frac{x^2}{\sqrt{z^2-x^2}}\left[\frac{8z}{x^2}+\ln\left(\frac{2-z}{2+z}\right)\right]\nn.
\eearr
where $z=\hbar\omega/\mu$, $x=q/k_F$. 
In the $z>x$ region for $\omega \to 0$, the poles in triplet channel are solutions to 
$\Re\chi^{(0)}(\vq, \omega)=1/U$. For $q\to 0$ we have,
\bearr 
   &&z={\frac {2 \left( 2\pi {v_F}^{2}+g\mu U \right) x}{\sqrt {
   \pi {v_F}^{2}g\mu U+16{\pi }^{2}{v_F}^{4}+{g}^{2}{\mu}^{2}{U}^{2}{x}^{2}}}}  \nn \\
   && \simeq \frac{1+g\mu U/2\pi v_F^2}{\sqrt{1+g \mu U/\pi v^2_F}}
   \left(x-\frac{(g\mu U/2\pi v^2_F)^2}{2(1+g\mu U/\pi v^2_F)}x^3\right)\nn \\
   &&\simeq\left(1+\gamma\right) \left(x-\gamma x^3\right),
   ~~~~~~~~\gamma=\frac{1}{2}(\frac{g\mu U}{2\pi v_F^2})^2,
   \label{s1-doped.eqn}
\eearr
where in the last step we have expanded around $\mu=0$ limit, and 
$\gamma$ is small positive constant.
In Figure~\ref{root.fig} we have plotted the dispersion relation~(\ref{s1-doped.eqn})
for $\gamma=0.1$. For $\mu\to 0$ additional slope $\gamma$ tends to zero, and the 
spin-1 collective mode branch becomes tangent to the PHC edge. 
Note that the whole scale in this figure is proportional to $\mu$. Therefore
in $\mu\to 0$ limit we will have a situation schematically shown in Figure~\ref{PHC.fig}(e),
with triangular window becoming smaller and the spin-1 collective mode coming closer to
the intra-band PHC. In doped graphene spin-1 collective mode for small momenta, 
decays into intra-band part of PHC in $\mu\to 0$ limit and it becomes both 
numerically and experimentally difficult to capture the collective mode.

To demonstrate the importance of overlap factors removing
the triplet branch of excitations from the triangular empty
window around the $\Gamma$ point, in appendix, we provide the
contour plot for the triplet RPA susceptibility without taking
the overlap factors into account. As can be seen clearly in
Fig.~\ref{u-change.fig}, if the overlap factors are ignored in
doped case, one will erroneously expect a flat branch of triplet
excitations for small momenta inside the triangular region.
However, it is the effect of wave-function overlaps as discussed
above that pushes the branch down to make it tangent to the 
continuum of intra-band particle-hole pairs. These discussions
based on the chirality of the states in the Dirac cone hold for 
the low-momentum part of the spectrum. However, for larger momentum 
transfer, as can be seen in Fig.~\ref{PHC.fig} the empty region
below the PHC is quite large and there would be no contribution
from the intra-band parts to possibly interfere with the triplet
branch. 

\begin{figure}[h]
\begin{center}
\includegraphics[width=8cm,height=7.2cm,angle=0]{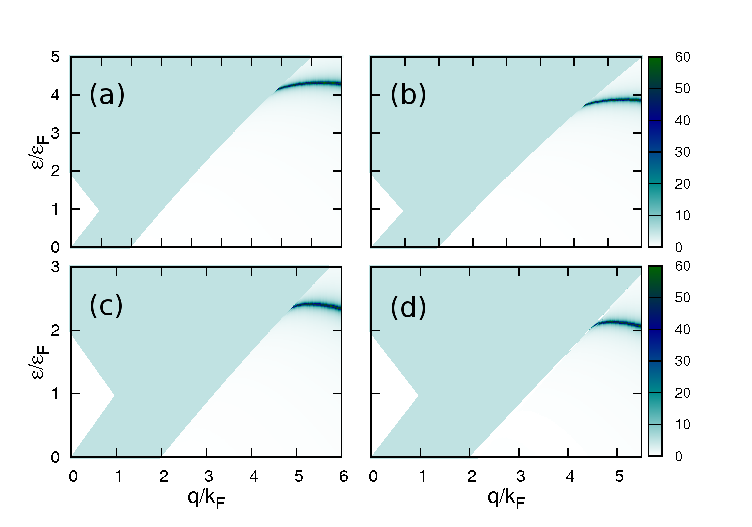}
\vspace{-4mm}
\caption{(Color online) Contour plot for $\Re\chi^{\rm RPA}_{\rm triplet}$ in $\Gamma \rightarrow M$ 
direction for doped graphene, panels correspond to (a) $\mu=0.4$ eV, $U=1.8t$, (b)
$\mu=0.4$ eV, $U=2.0t$, (c) $\mu=0.6$ eV, $U=2.0t$, (d) $\mu=0.6$ eV, $U=2.2t$.
}   
\label{doped-overlap.fig}
\end{center}
\vspace{-4mm}
\end{figure}

   Therefore, at finite $\mu$ and for the tight-binding band dispersion, we use
numerical integration to calculate the poles of $\chi^{\rm RPA}_{\rm triplet}$
at arbitrary momentum transfer.
This has been shown in Figure~\ref{doped-overlap.fig}. As can be seen,
in agreement with the phase space argument based on the overlap factors,
the small momentum part of the triplet collective mode is entirely lost
and nothing can be captured in the numeric data in the small 
triangular window. However, the collective mode emerges again 
in the larger window below the whole PHC, when one goes to
higher momentum transfers.
Panels (a) and (b) in Figure~\ref{doped-overlap.fig} correspond to
$\mu=0.4$ eV, with $U=1.8t$, $U=2.0t$, respectively. 
Panels (c) and (d) in this figure correspond
to $\mu=0.6$ eV, and $U=2t$, $U=2.2t$, respectively.  As can be seen,
for a given value of $\mu$, larger values of $U$ push the spin-1 collective mode 
to lower energies and give rise to larger binding energies. This feature is similar to the
case of undoped graphene. Comparing panels (b) and (c) which correspond
to the same value of $U=2t$ with different chemical potentials, one can see
that in the case of smaller $\mu=0.4$ eV, the energy of spin-1 collective mode
is on the scale of $\sim 4\mu=1.6$ eV. For $\mu=0.6$ eV, the energy of the
collective mode will be $\sim 2.5\mu=2.5\times(0.6)=1.5$ eV, which is not much
different from the energy scales shown in Figure~\ref{undoped-overlap.fig}. 

\subsection{Role of the other Dirac cone}
Our analytical results so far are valid in the long-wavelength limit where $\vec q$ is 
restricted to be around the $\Gamma$ point. Here we study our collective mode 
for momentum transfer near corners of the BZ $(\vq\approx\vec K_{i})$ in Figure~\ref{BZ.fig}.
This subsection is based on Ref.~\cite{Mikhailov}. To be self-contained, we 
quote their results specializing to triplet channel. It turns out that
singlet (plasmon) and triplet excitations around $\vec K_i$ points will be degenerate
for non-zero doping. 
The dielectric function in the triplet channel can be written as~\cite{Mikhailov},
\bearr
   \epsilon_{\vec Q,\vec Q'}(\vq,\omega)=\delta_{\vec Q,\vec Q'}+U \chi^{(0)}_{\vec Q,\vec Q'}(\vq,\omega),
   \label{dielectric}
\eearr
Note the difference between the signs of the interaction matrix elements~\cite{BickersScalapino} 
in the above equation with that of Ref.~\cite{Mikhailov}. 
The bare particle-hole bubble acquires matrix structure with respect to 
vectors connecting cones as,
\bearr
   &&\chi^{(0)}_{\vec Q,\vec Q'}(\vq, \omega) =\nn\\
   && \frac{g_{s}}{\sqrt{N}}\sum_{\vk} \sum_{\alpha,\alpha'}
   \frac{f^{\alpha}_{\vk}-f^{\alpha'}_{\vk +\vq}}{\hbar \omega +
   \epsilon_{\vk,\alpha} - \epsilon_{\vk+\vq,\alpha'}+ i0^{+}} \eta_{\vq, \vec Q}\eta^{*}_{\vq, \vec Q'},\ \ \
   \label{inter-vall-polar}
\eearr
where $\vec Q,\vec Q'=\{\vec 0,\vec K_1,\vec K_2\}$ are shown in figure~\ref{BZ.fig} 
and $\epsilon_{\vec Q,\vec Q'}$ is 3$\times$3 acquires a tensor structure with respect 
to the above indices. Here $\alpha,\alpha'$ refer to conduction (+) and valence (-) bands. 
Overlap factors in this case as given in Ref.~\cite{Mikhailov} become,
\be
   \eta_{\vq, \vec Q}=\frac{1}{2}M(|\vq+\vec Q|)
   \left[\zeta_{k}\zeta^{*}_{k+q}+\alpha \alpha'e^{-i(\vq+\vec Q).\vec a}\right],
\ee
where $\vec a$ is a basis vector connecting two carbons on the same 
sub-lattice of the honeycomb lattice and the atomic form factor is
$M(|\vq+\vec Q|)=\int_{3D}{d^3r |\psi(\vec r)|^2 e^{i(\vq+\vec Q).\vec a}}$,
with $\psi(\vec r)$ being atomic $p_z$ orbital and integration is performed 
in 3D space. Poles of the tensor susceptibility are given by the following
condition~\cite{Mikhailov},
\begin{equation}
    \det\left[\epsilon_{\vec Q,\vec Q'}(\vq,\omega)\right]= 0,
   \label{det}
\end{equation}
where $\vec q$ is around $\vec K_i$ and the dielectric tensor is as given
in Ref.~\cite{Mikhailov}, 
\begin{equation}
   \epsilon_{\vec Q,\vec Q'}=\delta_{\vec Q,\vec Q'}
   +\frac{\beta}{\sqrt{|\omega|/v_{F}q-1}}f_{\vec Q}(\theta)f^{*}_{\vec Q'}(\theta)
   \label{dielectric.eqn}
\end{equation}
with $f_{\vec Q}(\theta)=(-e^{-2i\theta}+e^{-i\vec Q.\vec a})/\sqrt{6}$ 
and $\theta$ is angle between $e. g.$ the vector  $\vq - \vec K_1$ and 
its neighboring non-equivalent cone $\vec K_{2}$ in Figure~\ref{BZ.fig}. 
Solution of Eq.~(\ref{det}) is given by (note that the sign of $\beta$ in 
Eq.~(\ref{dielectric.eqn}) for triplet channel is different from singlet 
channel),
\begin{equation}
   \omega_{\rm spin}(\vq)=v_{F}(1+\beta^2)|\vq-\vec K_{i}|=v_{s}|\vq-\vec K_{i}|.
   \label{spinK.eqn}
\end{equation}
where $v_{s}=v_{F}(1+\beta^2)$ is  velocity for spin mode and 
$|\vq-\vec K_{i}|a \ll  1$, $\beta=\frac{3g_{s}U \mu}{4\pi\sqrt{2} v^{2}_{F}}M^{2}(K)$.
Therefore the spin collective mode, near $\vec K_i$ points 
has linear dispersion with a slope slightly higher than the PHC edge. 
In $\mu\to 0$ limit it reduces to earlier findings of 
Ref.~\cite{BaskaranJafari}. The sign of interaction which is encoded
in parameter $\beta$ is irrelevant in the above discussion. Therefore
near $\vec K_i$ points in doped graphene singlet (plasmon) and triplet
collective modes are degenerate. However, the difference shows up in the
undoped graphene corresponding to $\mu=0$, where there would be no 
room for singlet (plasmon) collective modes below the PHC, and the repulsion
from inter-band  particle-hole states stabilizes the triplet collective mode
by a $|\vec q-\vec K_i|^3$ correction~\cite{BaskaranJafari}.

\section{Summary and discussion}
We investigated dispersion of a triplet neutral collective mode
in graphene. We revisited the problem of undoped graphene, and
showed that in the case of undoped graphene, inclusion of wave function
overlap factors does not qualitatively modify the dispersion of 
neutral spin-1 collective mode formed as a split-off state below the
PHC. 
%Therefore the comment concerning matrix character of susceptibility 
%in real space~\cite{PeresComment} is irrelevant. 
In the case of doped graphene, PHC acquires additional intra-band
portion, while at the same time some portions of the inter-band PHC
will be lost due to Pauli blocking. In this case there will be small
triangular windows adjacent to energy axis near $\Gamma$ and $\vec K_i$
corners of hexagonal BZ. Near $\Gamma$ point, the intra-band overlap 
factors at small momentum transfers approach zero, which in turn
shrinks the phase space required for the formation of triplet collective excitation,
thereby pushing the triplet branch to be tangent to the continuum of intra-band
excitations. 
The neutral spin-1 branch will emerge again below the total PHC at larger
momentum transfers. Very close to $\vec K_i$ corners of the hexagonal BZ, 
singlet (plasmon) collective mode will be degenerate with the triplet
excitations dispersing linearly~\cite{Mikhailov}. 
Neutron scattering signals involving spin flip can serve as a probe to 
isolate the contribution of near $K_i$ triplet excitations from degenerate
singlet plasmon excitation. 

   Recent large scale projective quantum Monte Carlo calculation
indicates presence of substantial spin liquid character in the regime of 
intermediate correlation on honeycomb lattice at half-filling~\cite{Meng}. 
Our diffusion Monte Carlo (DMC) study of $sp^2$ bonded systems, 
as well as exact diagonalization study of the Hubbard model on 
honeycomb geometry supports a triplet collective excited states in 
these systems. Such triplet states exist even in the limit of strong
correlations on the honeycomb lattice~\cite{Noorbakhsh,Mosadeq}.

\section{Acknowledgements}
We thank K. Haghighi for technical assistance in computing facilities. 
S.A.J. was supported by the Vice Chancellor for Research Affairs of 
the Isfahan University of Technology, and the National Elite Foundation (NEF) of Iran.

\section{Appendix}

\begin{figure}[h]
\begin{center}
\includegraphics[width=9cm,height=8cm,angle=0]{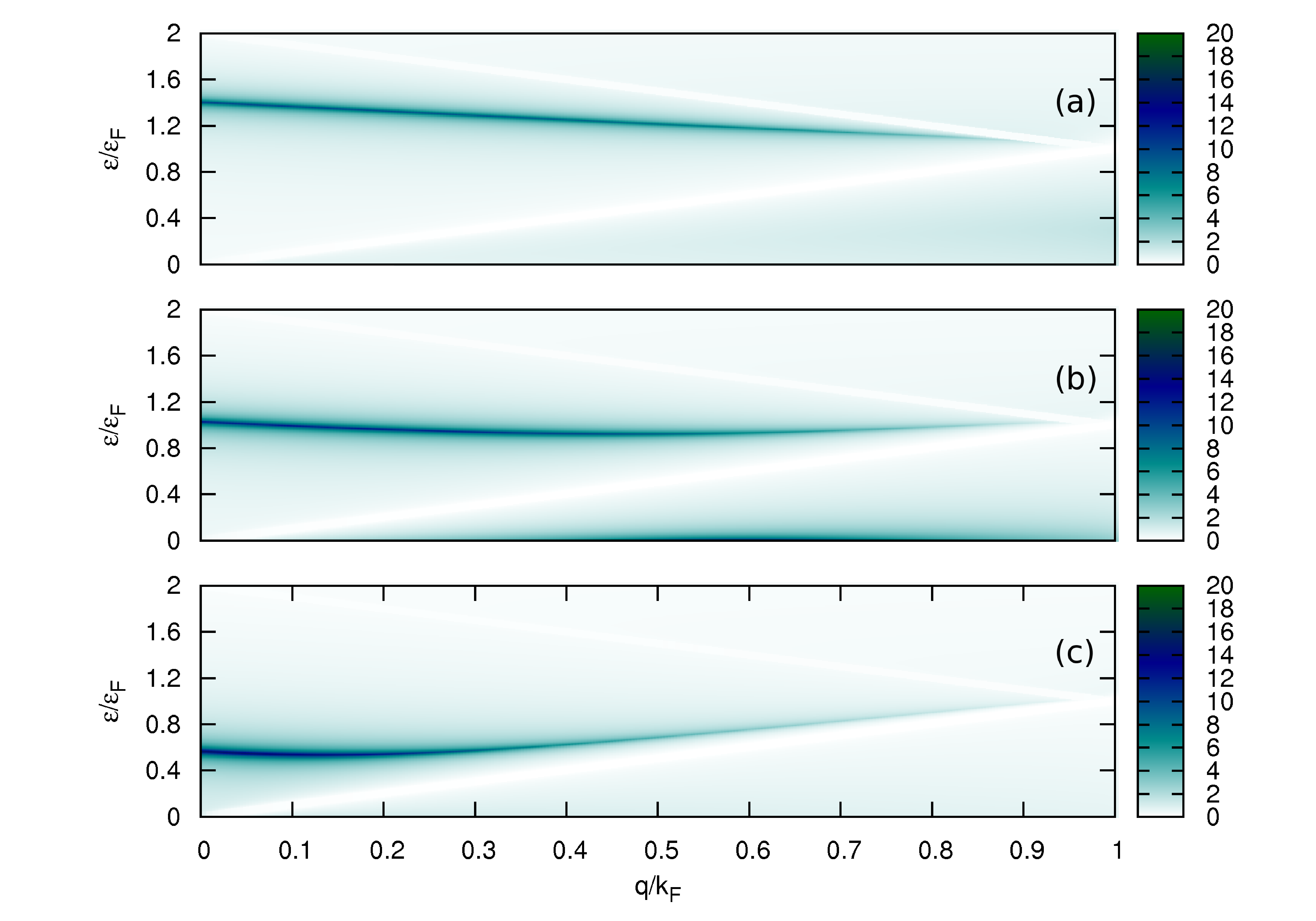}
\vspace{-5mm}
\caption{(Color online)  Numerical calculation of the spin-1 
collective mode for (a) U=1.8t, (b) U=2t, (c) U=2.2t and $\mu=0.4$ eV.
Borders of the triangular region of Fig.~\ref{PHC.fig} (e) are shown 
in white here.
}
\label{u-change.fig}
\end{center}
\vspace{-4mm}
\end{figure}

In this appendix we calculate the triplet susceptibility
without taking into account the overlap-factors in doped
graphene. This has been done both numerically and analytically.
This demonstrates that unlike the case of undoped graphene,
where it was not essential to include these factors, in doped
graphene, they give rise to a flat dispersion for the triplet
collective excitations. The proper inclusion of the overlap factors
as discussed in the text pushes such a dispersion down to touch
the upper border of intra-band PHC.
The numerical evaluation of the triplet susceptibility by
dropping the overlap factors over the
whole BZ has been presented in Fig.~\ref{u-change.fig}.

Now we obtain analytic expressions for the total susceptibility in
doped graphene by ignoring the overlap factors. There are two sets
of intra-band and inter-band terms. In the following, we calculate them separately.

\subsection{Intra-band term}
The nature of PHC is shown in Figure~\ref{PHC.fig} (e). Part of PHC 
shown in blue corresponds to intra-band particle-hole processes.
As can be seen this part is qualitatively similar to the PHC of 
standard 2D metals with extended Fermi surface. The only difference
is in the form of dispersion which unlike ordinary metals with quadratic
dispersion, here one has cone like dispersion.
For intra-band particle-hole excitations we have,
\be
\chi^{(0)}_{\rm intra}(\vec{q},\omega) =\frac{1}{N} \sum_{\vec{k}} \frac{f_{c,\vec{k}+\vec{q}}-f_{c,\vec{k}}}
{\hbar\omega-(\epsilon_{{\vec{k}+\vec{q}}}-\epsilon_{\vec{k}})+i0^{+}}.
\label{intra.eq}
\ee
In the $T\to 0$ limit where,
\bearr
f_{c,\vec{k}+\vec{q}} &=& \frac{1}{e^{\beta(\epsilon_{\vec{k}+\vec{q}}-\mu)}+1} 
\rightarrow \Theta(\mu-\epsilon_{\vec{k}+\vec{q}}), \label{step1.eq} \\
f_{c,\vec{k}} &=& \frac{1}{e^{\beta(\epsilon_{\vec{k}}-\mu)}+1} 
\rightarrow \Theta(\mu-\epsilon_{\vec{k}}) \label{step2.eq},
\eearr
it becomes,
\be
\chi^{(0)}_{\rm intra}(\vec{q},\omega)=\frac{1}{N} \sum_{\vec{k}} 
\frac{\Theta(\mu-\epsilon_{\vec{k}+\vec{q}})-\Theta(\mu-\epsilon_{\vec{k}})}
{\hbar\omega-(\epsilon_{{\vec{k}+\vec{q}}}-\epsilon_{\vec{k}})+i0^{+}}.
\ee 
In the first step function, we apply a change of variable,
$\vec{k}+\vec{q} \rightarrow -\vec{k}$, 
and after converting the summation to integral, we obtain:
\bearr  
&&\chi^{(0)}_{\rm intra}(\vec{q},\omega)=\frac{A}{4\pi^{2}} \int dk^2 
( \frac{\Theta(k_{F}-k)}{\hbar\omega-(\epsilon_{\vec{k}}-\epsilon_{{\vec{k}+\vec{q}}})+i 0^{+}}    \nn\\
&&-\frac{\Theta(k_{F}-k)}{\hbar \omega-(\epsilon_{{\vec{k}+\vec{q}}}-\epsilon_{\vec{k}})+i 0^{+}}),
\label{chiintra.eqn}
\eearr 
where $A$ is the unit cell area.
In analytic calculations we use $\epsilon_{\vec k}=\hbar v_F|\vec k|$
which is valid for low energies. For arbitrary $\vec q$ we evaluate the integrals with numerical
quadratures.
Using the formula $\frac{1}{x+i 0^+}=\textit{P}\frac{1}{x}-i\pi\delta(x)$, 
the imaginary part can be most conveniently written as,
\bearr
&&\Im\chi^{(0)}_{\rm intra}({\vec{q},\omega})=\nn 
\\
&&-\frac{A}{4\pi^2}\int d^2\vec k \left[\delta(\hbar\omega-\epsilon_{\vec{k}}+\epsilon_{\vec{k}+\vec{q}})-
\delta(\hbar\omega+\epsilon_{\vec{k}}-\epsilon_{\vec{k}+\vec{q}})\right]=\nn 
\\
&& -\frac{A}{4\pi^2}\int_{0}^{k_{F}}\!\!\!\! kdk \int_{0}^{2\pi}\!\!\!\!\!d\phi [\delta(z-k+\sqrt{k^2+q^2+2kq\cos\phi})\nn 
\\ 
&&-\delta(z+k-\sqrt{k^2+q^2+2kq\cos\phi})],
\label{a1.eqn}
\eearr
where $z=\omega/v_{F}$. Delta integrals are simplified by using the 
formula $\delta(f(x))=\sum_{s}\frac{\delta(x-s)}{|\nabla_{x}f(x)|_{x=s}}$ 
where $s$ denotes a root of $f(x)$. Let us define,  
\be 
f(k,q,\phi)=z\pm(k-\sqrt{k^2+q^2+2kq\cos\phi}).
\ee 
In term of new variable $u=\cos \phi$, the root of 
$f(k,q,\phi)=0$ is, $u=\frac{z^{2}-q^2 \pm 2zq}{2kq}\nn$,
which gives, 
\be
\nabla_{u}f(k,q,u)=-\frac{kq}{z\pm k}.
\ee
Substituting,
\be
d\phi=-\frac{du}{\sqrt{1-u^2}}, 
\ee
in Eq.~(\ref{a1.eqn}) the $u$ integral becomes trivial and we
are left with the following integration over radial variable $k$:
\bearr
&&\Im\chi_{\rm intra}^{(0)}(\vec{q},z)=\frac{-A}{2\pi \hbar v_{F}}\int_{0}^{k_{F}}dk \times  \\
&&\left(
\frac{(k-z)\Theta(k-\frac{z+q}{2}) }{q\sqrt{1-(\frac{z^2-q^2-2zk}{2kq})^2}} 
-\frac{(k+z)\Theta(k-\frac{q-z}{2})}{q\sqrt{1-(\frac{z^2-q^2+2zk}{2kq})^{2}}}
\right)\nn
\eearr

The step functions in integrand correspond to particle-hole (p-h) 
continuum. We restrict ourselves to $q<2k_{F}$ region. In this region,
the dissipative part of intra-band processes is non-zero only when 
$\omega<qv_{F}$. Radial integration can be performed to give~\cite{Abramowitz,Wunsch},

\bearr 
&&\Im\chi_{\rm intra}^{(0)}(\vec{q},z)=\frac{-A}{16\pi \hbar v_{F}}\frac{q}{\sqrt{1-z^2/q^2}}\times\nn\\  
&&\left[\left(1-2\frac{z^2}{q^2}\right)\ln\left(\left(\frac{2k_{F}-z}{q}\right)+
\sqrt{\left(\frac{2k_{F}-z}{q}\right)^2-1}\right) \right. \nn\\
&&+\left(\frac{2k_{F}-z}{q}\right)\sqrt{\left(\frac{2k_{F}-z}{q}\right)^2-1} \\
&&\left.-\left(1-2\frac{z^2}{q^2}\right)\ln\left(\left(\frac{2k_{F}+z}{q}\right)+\sqrt{\left(\frac{2k_{F}+z}{q}\right)^2-1}\right) \right.\nn\\
&&\left.-\left(\frac{2k_{F}+z}{q}\right)\sqrt{\left(\frac{2k_{F}+z}{q}\right)^2-1}\right]\nn
\label{imxintra.eqn}
\eearr

For calculation of $\Re\chi^{(0)}_{\rm intra}(\vec{q},\omega)$ , 
we directly use Eq.~(\ref{intra.eq}) and we find:
\bearr 
&&\Re\chi^{(0)}_{\rm intra}(\vec{q},z)=\frac{A}{4\pi^{2}\hbar v_{F}}\int d^2\vec k \nn\\
&&\left[\frac{1}{z-k+\sqrt{k^2+q^2+2kq\cos\phi}} \right. \nn\\
&&\left. -\frac{1}{z+k-\sqrt{k^2+q^2+2kq\cos\phi}}\right]\nn \\
&&=\frac{A}{2\pi^{2}\hbar v_{F}}\int_{0}^{k_{F}}kdk \int_{-\pi}^{\pi}d\phi \times\nn\\
&& \frac{k-\sqrt{k^2+q^2-2kq\cos\phi}}{z^2-(k-\sqrt{k^2+q^2-2kq\cos\phi})^2}.
\eearr  
 
The $\phi$ integral can be evaluated and simplified to give,
\bearr
&&\Re\chi^{(0)}_{\rm intra}(\vec{q},z)|_{z>q}= 
\frac{A}{\pi \hbar v_{F}}\int_{0}^{k_{F}} dk\times \nn\\
&&\left(\frac{k(z-k)}{\sqrt{[(q-k)^2+(z-k)^2][(q+k)^2-(z-k)^2]}}\right. \nn\\
&&\left. -\frac{k(z+k)}{\sqrt{[(z+k)^2-(q-k)^2][(z+k)^2-(q+k)^2]}}\right)\nn \\
&&= \frac{A}{16\pi \hbar v_{F}}\frac{1}{\sqrt{z^2-q^2}}\times\nn\\ 
&&\left[(q^2-2z^2)\ln\left((\frac{2k_{F}+z}{q})+\sqrt{(\frac{2k_{F}+z}{q})^2-1}\right)\right.\nn \\
&&\left. -(q^2-2z^2)\ln\left((\frac{2k_{F}-z}{q})+\sqrt{(\frac{2k_{F}-z}{q})^2-1}\right)\right.\nn\\
&&\left. -(q^2-2z^2)\ln\left((\frac{z}{q})+\sqrt{(\frac{z}{q})^2-1}\right)\right.\nn\\
&&\left.-(q^2-2z^2)\ln\left(-(\frac{z}{q})+\sqrt{(\frac{z}{q})^2-1}\right)\right.\nn\\
&&\left. -q^2(\frac{2k_{F}-z}{q})\sqrt{(\frac{2k_{F}-z}{q})^2-1}\right. \nn\\
&&\left. +q^2(\frac{2k_{F}+z}{q})\sqrt{(\frac{2k_{F}+z}{q})^2-1}\right].
\label{rexintra.eqn}
\eearr
Demanding the expressions under square root to be positive, 
gives the following region for the window protected from free 
particle-hole energy levels: $q<z<-q+2k_{F}$.

\subsection{Inter-band term}
Inter-band processes  (near $\Gamma$ and $K$ points) one has,
\be
\chi^{(0)}_{\rm inter}(\vec{q},\omega) =\frac{1}{N} \sum_{\vec{k}} \frac{f_{c,\vec{k}+\vec{q}}-f_{v,\vec{k}}}
{\hbar\omega-(\epsilon_{{\vec{k}+\vec{q}}}+\epsilon_{\vec{k}})+i0^{+}},
\label{inter.eq}
\ee
which in an analogous way to Eq.~(\ref{chiintra.eqn}) simplifies to,
\bearr  
&&\chi^{(0)}_{\rm inter}(\vec{q},\omega)=\frac{A}{4\pi^{2}} 
\int d^2\vec k \times \\
&&\left( \frac{\Theta(k_{F}-k)}{\hbar\omega-(\epsilon_{\vec{k}}+\epsilon_{{\vec{k}+\vec{q}}})+i 0^{+}}  
-\frac{1}{ \hbar\omega-(\epsilon_{{\vec{k}+\vec{q}}}+\epsilon_{\vec{k}})+i 0^{+}}\right).\nn
\eearr 
When working with the linearized
low-energy theory, the limits are from $0$ to a momentum 
cutoff $k_c$ of the linearized theory.
Substituting linear dispersion in polar coordinates, 
the inter-band part simplifies to,
\bearr
&&\chi^{(0)}_{\rm inter}(\vec{q},\omega)=\nn \frac{-A}{4\pi^{2}} 
\int_{k_{F}}^{k_{c}} kdk \int _{0}^{2\pi} d\phi \times \\
&&\frac{1}{\hbar \omega-\hbar v_{F}(k+\sqrt{k^2+q^2+2kq\cos\phi})+i 0^{+}}.
\label{chiinter.eqn}
\eearr

The imaginary part of $\chi^{(0)}_{\rm inter}$ can be written as,
\bearr
&&\Im\chi^{(0)}_{\rm inter}(\vec{q},z)=\frac{A}{4\pi v_{F}} \int_{k_{F}}^{k_{c}} kdk \times \nn\\
&&\int _{0}^{2\pi}\delta\left(z-k-\sqrt{k^2+q^2+2kq\cos\phi}\right).
\eearr
First we do integration on $\phi$, to find~\cite{JafariBaskaran}:
\bearr
&&\Im\chi_{inter}^{(0)}(\vec{q},z)=\frac{A}{2\pi \hbar v_{F}}\int_{k_{F}}^{k_{c}}dk\frac{z-k}{q\sqrt{1-(\frac{z^2-q^2-2zk}{2kq})^2}}\nn\\
&&\times \left[\Theta(k-\frac{z+q}{2})  -\Theta(\frac{z-q}{2}-k)\right]\nn \\
&=&\frac{A}{16\hbar v_{F}}\frac{2z^2-q^2}{\sqrt{z^2-q^2}}, -q+2k_{F}<z< q+2k_{c}.
\label{imxinter.eqn}
\eearr
Now we use Kramers-Kronig relation for calculation of $\Re\chi^{(0)}_{\rm inter}$ 
from imaginary part, $\chi^{(0)}_{\rm inter}$~\cite{JafariBaskaran}:
\be
\Re\chi^{(0)}_{\rm inter}(\vec{q},\omega) =
 \frac{A}{16\pi \hbar v^{2}_{F}}\!\int^{(q+2k_{c})v_{F}}_{(-q+2k_{F})v_{F}}
\frac{d\omega'}{\omega'-\omega}\frac{2\omega^{2}-q^{2}v^{2}_{F}}{\sqrt{\omega^{2}-q^{2}v^{2}_{F}}}.
\ee 
Defining the new variable $\eta$ by relation $\omega'=qv_{F}\coth(\eta)$, 
the limits of integration $\eta_1,\eta_2$ are given by 
$\coth(\eta_{2})=1+2k_{c}/q$, $\coth(\eta_{1})=-1+2k_{F}/q$, so that
we obtain:
\bearr
&& \Re\chi^{(0)}_{\rm inter}(\vec{q},\omega)=
\frac{Aq^2}{16\pi\hbar}\int^{\eta_{2}}_{\eta_{1}}d\eta \frac{2\coth^{2}(\eta)-1}{\sinh(\eta)\left[\omega-qv_{F}\coth(\eta)\right]}\nn\\
&&=\frac{A}{16 \pi\hbar}\left\{\frac{-q}{v_{F}}\left(2+2\frac{k_c-k_F}{q}+\frac{2k_F}{q}
\sqrt{1-\frac{q}{k_{F}}}-\frac{2k_c}{q}\sqrt{1+\frac{q}{k_{c}}}\right)\right.\nn\\
&&\left.-\frac{2\omega}{v^{2}_{F}}\ln\left(\frac{-1+\frac{2k_{F}}{q}-\frac{2k_{F}}{q}\sqrt{1-\frac{q}{k_F}}}
{1+\frac{2k_{c}}{q}-\frac{2k_{c}}{q}\sqrt{1+\frac{q}{k_c}}}\right)\right. \nn \\
&&+\frac{2q^2(1-2\omega^2/v_F^2)}{\sqrt{q^{2}v^{2}_{F}-\omega^2}}\left[
\arctan\left(\frac{qv_F(1+\frac{2k_{c}}{q}-\frac{2k_{c}}{q}\sqrt{1+\frac{q}{k_c}})-\omega}{\sqrt{q^{2}v^{2}_{F}-\omega^2}}\right)\right.\nn\\
&&\left. -\arctan\left(\frac{qv_F(-1+\frac{2k_{F}}{q}-\frac{2k_{F}}{q}\sqrt{1+\frac{q}{k_F}})-\omega}{\sqrt{q^{2}v^{2}_{F}-\omega^2}}\right)\right]
\label{rexinter.eqn}\\
%&&-\frac{4q^{2}\omega^2}{v^{2}_{F}\sqrt{q^{2}v^{2}_{F}-\omega^2}} \left[
%   \arctan(\frac{qv_F(1+\frac{2k_{c}}{q}-\frac{2k_{c}}{q}\sqrt{1+\frac{q}{k_c}})-\omega}{\sqrt{q^{2}v^{2}_{F}-\omega^2}})
%   -\arctan(\frac{qv_F(-1+\frac{2k_{F}}{q}-\frac{2k_{F}}{q}\sqrt{1+\frac{q}{k_F}})-\omega}{\sqrt{q^{2}v^{2}_{F}-\omega^2}}) 
%\right]\nn\\
&&\left. +\frac{q}{v_{F}(1+\frac{2k_{c}}{q}-\frac{2k_{c}}{q}\sqrt{1+\frac{q}{k_c}})}
-\frac{q}{v_{F}(-1+\frac{2k_{F}}{q}-\frac{2k_{F}}{q}\sqrt{1+\frac{q}{k_F}})}\right\}.\nn
\eearr
Equations~(\ref{imxintra.eqn}), (\ref{rexintra.eqn}), (\ref{imxinter.eqn}) and (\ref{rexinter.eqn})
complete the analytic evaluation of total non-interacting susceptibility 
$\chi^{(0)}=\chi^{(0)}_{\rm intra}+\chi^{(0)}_{\rm inter}$.
The final results we use in our plots
are given by Equations~(\ref{imxinter.eqn}) and (\ref{rexinter.eqn})
which are again valid for values of $|\vec q|$ which are small compared
to $k_F$. These results agree with numerical results when plotted on the
same figure. As can be seen in Figure~\ref{u-change.fig}, neglecting the 
wave function overlap factors in doped graphene gives rise to a
a neutral triplet collective mode branch in small $q$ region, which 
has essentially no dispersion. This is in contrast to Figure~\ref{doped-overlap.fig},
where the inclusion of overlap factors essentially destroys the triplet
collective mode branch. Therefore we can see that, in contrast
to undoped graphene, where neglect of overlap factors did not lead to
any qualitative change in the fate of neutral triplet collective mode,
in doped graphene, overlap factors assume very essential role in the
case of doped graphene.

\end{document}